\documentclass[conference, a4paper]{IEEEtran}
\IEEEoverridecommandlockouts

\ifCLASSINFOpdf
  \usepackage[pdftex]{graphicx}
  \DeclareGraphicsExtensions{.pdf,.jpeg,.png}
\else
\fi

\usepackage{cite}
\usepackage{amsmath,amssymb,amsfonts}
\usepackage{amsthm}
\usepackage{mathtools}
\usepackage{algorithmic}
\usepackage{graphicx}
\usepackage{textcomp}
\usepackage[left=1.4cm, right=1.4cm, top=1.7cm]{geometry}
\usepackage{caption}
\usepackage{anyfontsize}
\usepackage{csquotes}
\usepackage{xcolor}
\usepackage{multirow}
\usepackage{tabularx}
\usepackage{algorithm}
\usepackage[algo2e]{algorithm2e}
\usepackage{hyperref}
\usepackage{makecell}

\usepackage{bm}
\usepackage{enumitem}
\usepackage{bbm}

\usepackage{longtable}
\usepackage{flushend}

\makeatletter
\newcommand*\titleheader[1]{\gdef\@titleheader{#1}}
\AtBeginDocument{%
\let\st@red@title\@title
\def\@title{%
\bgroup\normalfont\normalsize\centering\@titleheader\par\egroup
\vskip0.2em\st@red@title}
}
\makeatother

\makeatletter
\renewcommand{\fnum@figure}{Figure \thefigure}
\makeatother

\title{{The Development of the Reproductive Healthcare Equity Algorithm (RHEA)} \\ }

\titleheader{Andrew P. Sage Memorial Capstone Design Competition}

\author{\IEEEauthorblockN{Shriya Karam\textsuperscript{*}, Lauren Shanos, Jessica Ford, Lorenzo Castaneda, Megan S. Ryerson, Rakesh Vohra \\
University of Pennsylvania \\
Philadelphia, PA, USA\\ 
*Corresponding Author \\ 
\{\href{mailto:karams@seas.upenn.edu}{karams}, 
\href{mailto:lshanos@seas.upenn.edu}{lshanos},
\href{mailto:fordje@seas.upenn.edu}{fordje},
\href{mailto:lorca@seas.upenn.edu}{lorca},
\href{mailto:mryerson@design.upenn.edu}{mryerson},
\href{mailto:rvohra@design.upenn.edu}{rvohra}\}@upenn.edu  }
}

\IEEEaftertitletext{\vspace{-1\baselineskip}}

\begin{document}

\maketitle
\thispagestyle{plain}
\pagestyle{plain}

\noindent \begin{abstract}
After the repeal of Roe vs. Wade in June 2022, women face long-distance travel across state lines to access abortion care. For women who also face socioeconomic hardship, travel for abortion care is a significant burden. To ease this burden, abortion access nonprofits are funding and/or supplying transportation to abortion clinics. However, due to the uneven distribution of demand and supply for abortions, these nonprofits don’t have efficient logistical operations. As a result, low-income, underserved women may not have access to adequate reproductive healthcare, thus widening healthcare inequity gaps. Nonprofits may also risk not serving the needs of vulnerable women without access to adequate reproductive healthcare, and in doing so, waste resources, money, and volunteer hours. To address these challenges, we create an interactive, web-based planning tool, the Reproductive Healthcare Equity Algorithm (RHEA), to guide nonprofits in strategically allocating resources and serving demand. RHEA leverages an optimization model to determine the maximum flow and minimum transportation cost to route women across a network of counties and abortion clinics, subject to transportation supply, budget, and time constraints for one day of operations for a nonprofit. In doing so, we collaborate with abortion access nonprofits to cater our model design and interface development to their needs and considerations. Ultimately, we seek to optimize resource allocation for nonprofits providing abortion care logistics and improve abortion access for low-income, underserved women.
\end{abstract}

\section{Introduction}
\label{sec:introduction}
The repeal of Roe vs. Wade in June 2022 has had a profound impact on access to reproductive health care. Women residing in states that have instituted abortion bans, which totals 14 states, now face a significant burden in traveling to access reproductive health care services \cite{guttmacher}. Moreover, for women who face other hardships, such as low-income status, single-parent households, or lack of vehicle access, travel for abortion care is a significant burden \cite{barr, jerman, henshaw}. To ease this burden, abortion access nonprofits are funding and/or supplying transportation to abortion clinics, in the form of volunteer drivers, pilots, and/or commercial aviation. However, due to the uneven distribution of demand and supply for abortions, these nonprofits don’t have efficient logistical operations. As a result, low-income, underserved women may not have access to adequate reproductive healthcare, thus widening healthcare inequity gaps. Nonprofits also risk not serving the needs of vulnerable women without access to healthcare, and in doing so, waste resources, money, and volunteer hours. 

Previous scholarship on reproductive health care access consists of modeling both demand- and supply-side abortion restrictions. While fertility control models provide understandings of how key factors and policy interventions affect observed abortion demand, abortion access nonprofits do not rely strictly on demand estimation of abortions. Given the dynamic (re)distribution of demand and supply for abortions after the repeal of Roe vs. Wade, abortion \emph{accessibility} is the key concern. However, quantitative methods in measuring abortion accessibility largely skew towards supply-based accessibility models that quantify the impact of travel distance to abortion clinics, which lack the nuances to consider individual-level effects in accessibility. Moreover, after the repeal of Roe vs. Wade, abortion accessibility (defined as travel distance) has decreased dramatically for women across the U.S., and as a result, abortion access nonprofit groups are concerned with providing logistical and travel support at a large scale to low-income, underserved women seeking abortion care. Thus, planning tools that determine the scale of resources for nonprofits needed to meet the anticipated demand for abortions is critical. Cost-effective transportation routes to in-person abortion care is an ever-pressing issue, especially with alternative forms of health care access facing legality and violence issues.

To address these health equity and logistical challenges, we design and implement a planning tool, the Reproductive Healthcare Equity Algorithm (RHEA), to guide nonprofits in their resource allocation to serve demand for abortions. RHEA is a combined optimization model and front end web interface system that allows nonprofits to understand how to most strategically allocate resources and ensure equitable demand satisfaction for abortions. Through an optimization model back end, RHEA allows the nonprofit user to input their resource estimates and constraints and understand the effect on total women transported, transportation cost, demand satisfaction, and transportation mode distributions. Thus the model determines the maximum flow and minimum transportation cost to route women throughout a network of counties and abortion clinics, subject to transportation supply, budget, and time constraints for one day of operations for a nonprofit. In implementing RHEA, we collaborate with abortion access nonprofits to cater our model design and interface development to their needs and considerations.

\section{Literature Review}
\label{sec:lit_review}
Abortion access is a heavily debated issue that has been litigated in the United States for several decades. Prior to the late 19th century, abortion was considered legal, but alongside the growth of the medical field and technologies, abortion became more restricted. Until the landmark case of Roe vs. Wade (1973), in which the Supreme Court ruled that access to abortion was a constitutional right, states largely regulated abortion as they saw fit. Even after this landmark decision, many rural states placed heavy restrictions on abortion access and continued to test the law in the following decades. For example, in the case of Planned Parenthood vs. Casey (1992), the Supreme Court allowed states to restrict abortion as long as it did not represent an undue burden on the person, in the form of mandatory waiting periods and counseling \cite{jhu}. After the case of Dobbs vs. Jackson Women's Health Organization (2022) ruled that states had the right to regulate abortion, many states quickly banned the medical procedure, leading to even more limited access to this essential service. 

\subsection{Demand-side Abortion Restrictions}
Before the repeal of Roe vs. Wade in 2022, research on reproductive health care and abortion access skewed heavily toward abortion demand estimation \cite{medoff1988, bearak}. Empirical analyses of the demand for abortions leverage fertility control models to estimate abortion rates at the state level as a function of variables such as the price of abortions, income, marital status, labor participation rate among women, religion, and whether the state had Medicaid funding \cite{medoff1988, garbacz}. Most notably, results show that increases in the price of abortion corresponds to a decline in abortion rate, while higher incomes correlate with a higher abortion rate. 

Existing research that empirically estimates demand for abortion provides an understanding of key factors that could influence abortion rates. However, these studies do not provide explanations as to how demand for abortions will be affected based on potential changes in restrictions. Several other studies use time series fertility control models to measure the effect on abortion rates due to abortion restrictions, such as parental involvement laws, accounting for abortion price and income factors \cite{medoff2007a, medoff2008a, joyce1997, joyce2006, myers_ladd}. These studies find that state mandatory waiting periods (which became common after Planned Parenthood vs. Casey) have no statistically significant impact on abortion rates, while parental involvement laws and Medicaid funding for abortions correlated with a decline in average state abortion rates.

While fertility control models and variations to account for year and abortion legislation provide explanations of the factors that explain abortion rates, demand estimation models may be limited as a whole in the modern climate of abortion restrictive laws in the U.S. Nonprofit organizations who seek to provide women with the resources they need to access reproductive health care do not strictly rely on research on demand estimations of abortions, according to our nonprofit partner Midwest Access Coalition \cite{mac}. Instead, the focus has significantly shifted towards providing patients with logistical and travel support to abortion clinics from an accessibility perspective. 

\subsection{Abortion Travel Burden and Accessibility}
Even before the repeal of Roe vs. Wade, literature on access to reproductive health care included analysis of accessibility to abortion and reproductive health care clinics. Due to abortion restrictions that vary by state in the U.S., women in places with restrictive abortion laws may take physical and financial risks to access abortion care \cite{bearak}. In the abortion care literature, accessibility is defined as the distance that women of reproductive age are to their nearest abortion clinics \cite{myers2021}. Across the U.S., travel distance to clinics varied significantly before the repeal of Roe vs. Wade: 39\% of U.S. women of reproductive age lived in a county with no abortion clinics, and more than 17\% of those obtaining an abortion in 2014 traveled 50 miles or more for care \cite{liebertpub}. 

Several approaches seek to determine the effect of travel distance and other barriers on abortion access. Economic models estimate the causal effect of distance to abortion facilities on abortion rates when modeling abortion rates as a Poisson model based on demographic factors at the county-level \cite{myers2021, thompson}. These studies find that increases in travel distance correlate with decreases in abortion rates \cite{myers2021, thompson, sethna}. Racial minorities, younger women, and rural women also face disproportionate impacts on travel distances to clinics \cite{myers2021, sethna}. Qualitative, survey-based studies also find that distance can be a significant travel burden and barrier to access abortion care for women \cite{sethna, jerman}. 

However, travel distance is not the sole factor to measure individuals’ burden in accessing abortion care. In reality, women face a suite of barriers that work together in complex interactions and exacerbate the already onerous experience of traveling for abortion care \cite{jerman, henshaw}. Beyond travel distance, factors that contribute to individual travel barriers can include: difficulty in financing abortion costs and travel, lack of insurance coverage, lack of information on abortion clinics, the cost of childcare services, and emotional burden due to stress and stigma \cite{barr, jerman, henshaw}. In this study, individual-level variables (e.g., emotional burdens, individual knowledge on abortion clinics) serve as key motivating factors in developing our planning tool to provide abortion access for underserved women. Furthermore, we move away from modeling access to abortion care based on travel time and distance exclusively, but also consider travel cost and key factors such as companionship and income as elements of abortion accessibility. 

\subsection{Post-Dobbs Abortion Access}
The repeal of Roe vs. Wade in 2022 led to significant supply-side restrictions due to clinic closures and restrictive laws across the U.S. \cite{guttmacher}. Travel times via surface transportation to the nearest abortion clinics were found to be statistically significantly higher at the median and mean after the Dobbs vs. Jackson decision. Additionally, the total percentage of women of reproductive age living more than 60 minutes from abortion facilities was found to be higher after the Dobbs decision \cite{rader}. Qualitative and quantitative findings from pre-Dobbs research show that low-income, rural, and racial minority women were more likely to bear a disproportionate impact of travel burden to abortion care; combining this with longer travel times after the Dobbs decision, these women are likely to face even more difficulties in accessing care. 

To address these issues, nonprofit groups are increasingly concerned with providing transportation and logistical support for women to abortion clinics based on the current distribution of supply and demand for abortions. Recent research has shown how nonprofit organizations are working around abortion restrictions to provide women with access to abortion care by transporting women via private pilots, volunteer drivers, commercial aircraft, or other modes of transportation \cite{politico}. For example, Elevated Access, a nonprofit organization, recruits volunteer pilots to transport women via private aircraft to abortion care to states with less abortion restrictions \cite{politico, rolling_stone}. Midwest Access Coalition, another abortion access nonprofit serving the Midwest region in the U.S., works with abortion patients to provide them with logistical and transportation support to abortion clinics.

Nonprofits providing transportation services face the challenge of sustaining their efforts at a large-scale. As of August 2022, Elevated Access has recruited 800 pilots but transported one woman \cite{politico}. In order to scale up their efforts, nonprofits require an understanding of the magnitude of resources, such as the monetary budget to purchase travel itineraries, volunteer pilots, and volunteer drivers required to service demand for abortions. To provide nonprofits with this knowledge, it is critical that they have planning tools to guide them on the scale of resources required to service demand, as well as an understanding of what additional resources are needed to ensure demand satisfaction.

\subsection{Abortion Pills and Telemedicine}
Before presenting the RHEA methodology, we first review the literature on alternative healthcare services to meet abortion demand. Alternative solutions to address inaccessible and inequitable abortion care include relying less on in-person care, and instead on telemedicine, mobile clinics, and abortion pills that may not require travel to abortion clinics. Scenario analysis to simulate low travel distances to abortion clinics for all counties shows that unmet need for abortion care can be captured through telemedicine \cite{thompson}. Abortion pills also provide safe and effective medications, with pills accounting for 54\% of all abortions in 2020 \cite{gutt_meds}. However, almost half of U.S. states are banning or placing heavy restrictions on abortion pills (i.e. mifepristone and misoprostol) via telemedicine, penalizing individuals who mail, dispense, or distribute abortion pills. Another alternative has been mobile abortion clinics that aim to increase travel distance to abortion care in rural areas. While mobile clinics could reduce travel times for women, and also alleviate the influx of demand for clinics in states with no abortion restrictions, these clinics can be the subject of vandalism and violence \cite{henshaw, doj}. As a result of the limitations of alternative forms of abortion care, low-cost and accessible transportation to in-person care still remains critical.

\section{Methodology}
\label{sec:methodology}
\begin{figure*}[!htbp]
\centering
\includegraphics[width=1.7\columnwidth]
{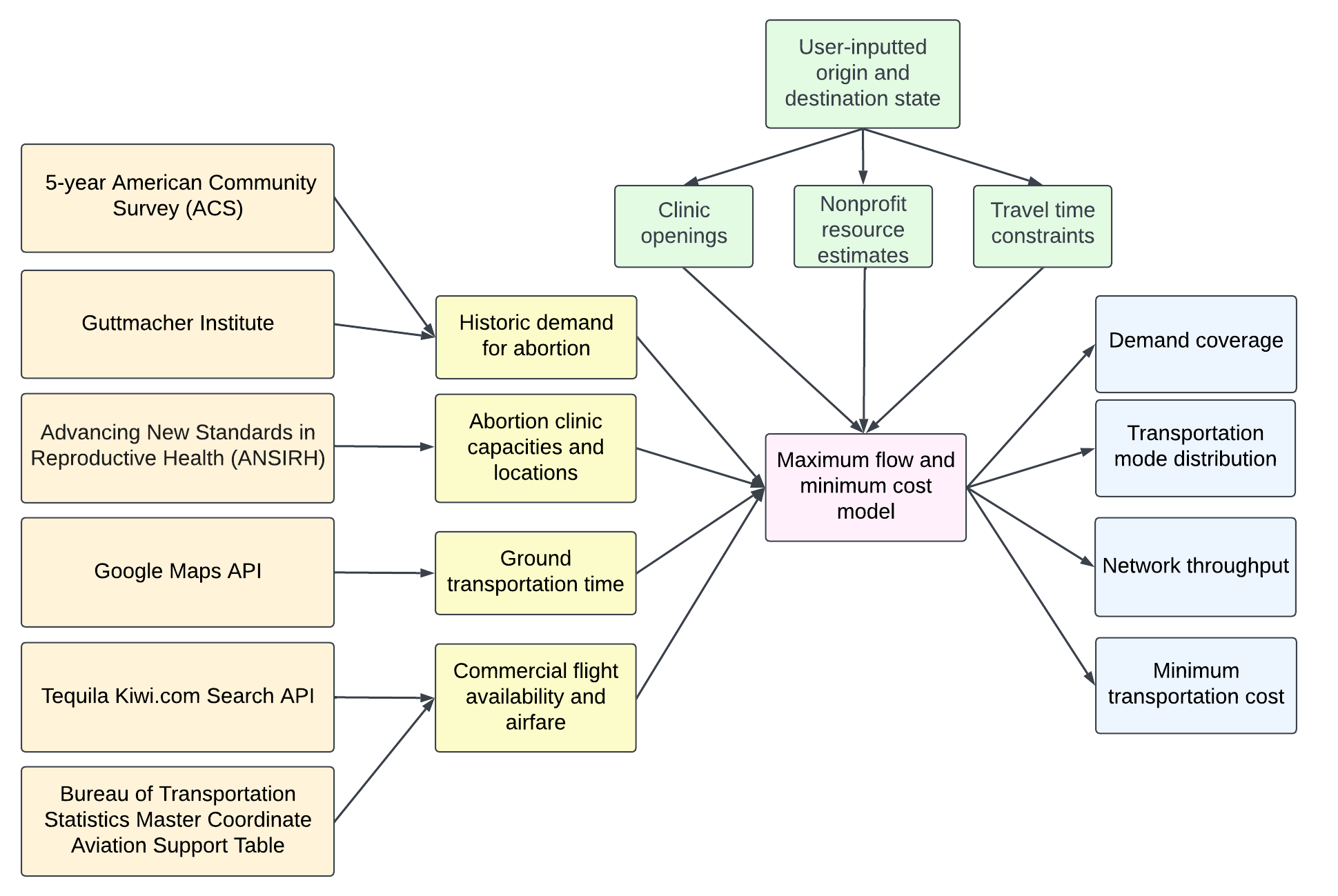}
\caption{Schematic of model inputs and outputs.}
\label{fig:chart1}
\end{figure*}

Figure \ref{fig:chart1} describes the overview of the RHEA methodology. The data sources and parameters listed in Figure \ref{fig:chart1} are inputs to the optimization model, which determine the maximum flow and minimum cost, transportation mode distribution, and demand satisfaction for routing women to abortion clinics. In the front end interface, the nonprofit user first must select the origin and destination state. Based on this selection, the user must then input their specific constraints as it pertains to budget and transportation supply (i.e. resource estimates), as well as travel time constraints and clinic openings, which are encoded into the optimization model as parameters and constraints. This interactive interface allows the user to vary input parameters and observe the effect on transportation cost, total network throughput, transportation mode distributions, and abortion demand satisfaction.

In this section, we first present a description of the model setup and notation (Section \ref{sec:model}). For the maximum flow model, we present the formulation, including the objective function and constraints, in Section \ref{sec:max_flow}. For the minimum cost model, we build on the formulation of the maximum flow model but modify the objective function and certain constraints in Section \ref{sec:min_cost}.

\subsection{Model Setup and Notation}
\label{sec:model}
The optimization models to determine maximum flow and minimum cost for routing women from abortion-restrictive states to abortion clinics are formulated as pure-integer linear programming problems. To set up the framework of our formulation, we consider a set of demand nodes (the spatial unit of analysis is the county) and supply nodes (abortion clinics), subject to capacity constraints across each arc considering one day's of operations for a nonprofit. Intermediate nodes include general and commercial airports, allowing individuals to be transported via air transportation as well as directly from their origin county to a destination clinic. In this model, we assume that each vehicle makes one trip in a single tour (i.e. no multiple stops). Women may be transported from their origin county to their destination clinics in one of the following three ways:
\begin{itemize}
    \item Drive from the origin county to a clinic, either by a volunteer driver or through ride hail (i.e. Uber/Lyft).
    \item Drive from the origin county to an origin commercial airport (either by a volunteer driver or through ride hail), fly to a destination commercial airport, and then drive from the destination airport to a clinic (either by a volunteer driver or through ride hail).
    \item Drive from the origin county to an origin general airport (either by a volunteer driver or through ride hail), fly to a destination general airport (via a volunteer pilot), and drive from the destination airport to the clinic (either by a volunteer driver or through ride hail).
\end{itemize}

In the following subsections, we describe the sets, parameters, and decision variables. 
\subsubsection{Sets}
\label{sec:sets}
The set notation throughout this model is:
\begin{itemize}
    \item $q \in Q$: set of origin counties 
    \item $m \in M$: set of origin commercial aviation airports
    \item $j \in J$: set of local access modes (includes ride hail and private vehicle)
    \item $g \in G$: set of origin general aviation airports
    \item $r \in R$: set of destination commercial aviation airports
    \item $p \in P$: set of destination general aviation airports
    \item $c \in C$: set of abortion clinics
    \item $v \in V$: set of counties that destination airports are in
\end{itemize}

\subsubsection{Parameters} The parameters used in this model are:
\label{sec:parameters}
\begin{itemize}
    \item $u$: volunteer pilots on standby in the origin state
    \item $z$: capacity of each general aviation aircraft
    \item $w$: capacity of each private vehicle
    \item $B$: total budget of the nonprofit for the day
    \item $\mathcal{A}$: maximum allowable airport access (i.e. county to airport) and egress (i.e. airport to clinic) travel time
    \item $\mathcal{F}$: maximum allowable in-flight time between origin and destination airports
    \item $T$: maximum allowable origin county to clinic travel time 
    \item $f_{m, r}$: commercial flight seat availability from $m$, the origin commercial airport, to $r$, the destination commercial across a day
    \item $a_{m, r}$: average airfare from $m$, the origin commercial airport, to $r$, the destination commercial airport across a day
    \item $\lambda_c$: number of individuals a clinic $c$ can service in a day
    \item $n_q$: supply of volunteer drivers in county $q$ in the origin state
    \item $n_v$: supply of volunteer drivers in destination county $v$ 
    \item $o$: ride hail cost per minute
    \item $p_q$: demand for abortion in county $q$ for one day
    \item $b_{p, v}$: binary variable denoting if destination general airport $p$ is in county $v$ (1 if yes, 0 otherwise)
    \item $b_{r, v}$: binary variable denoting if destination commercial airport $r$ is in county $v$ (1 if yes, 0 otherwise)
    \item $y$: binary variable denoting if all women to be routed in the origin state are traveling with a companion (1 if yes, 0 otherwise)
    \item $t_{q, g, j}$ travel time from county $q$ to origin general aviation airport $g$ on mode $j$
    \item $t_{q, m, j}$: travel time from county $q$ to origin commercial aviation airport $m$ on mode $j$
    \item $t_{g, p}$: travel time from origin general aviation airport $g$ to destination general aviation airport $p$ 
    \item $t_{m, r}$: travel time from origin commercial aviation airport $m$ to destination commercial aviation airport $r$
    \item $t_{r, c, j}$: travel time from destination commercial aviation airport $r$ to abortion clinic $c$ on mode $j$
    \item $t_{p, c, j}$: travel time from destination general aviation airport $p$ to abortion clinic $c$ on mode $j$
    \item $t_{q, c, j}$ travel time from origin county $q$ to abortion clinic $c$ on mode $j$
\end{itemize}

\subsubsection{Decision variables}
\label{sec:decision_variables}
The decision variables are all non-negative integer variables that represent the number of women traveling across the arcs in the network, including from origin counties, to general or commercial airports and by mode, and from destination general or commercial airports to clinics by mode. We also allow for individuals to be routed directly from origin counties to clinics across surface transportation modes.
\begin{itemize} 
\item $x_{q, g, j}$: flow from county $q$ to origin general aviation airport $g$  on mode $j$
\item $x_{q, m, j}$: flow from county $q$ to origin commercial aviation airport $m$ on mode $j$
\item $x_{q, g, p}$: flow from origin general aviation airport $g$ to destination general aviation airport $p$ by county $q$
\item $x_{q, m, r}$: flow from origin commercial aviation airport $m$ to destination commercial aviation airport $r$ by county $q$
\item $x_{q, r, c, j}$: flow from destination commercial aviation airport $r$ to abortion clinic $c$ by county $q$ on mode $j$
\item $x_{q, p, c, j}$: flow from destination general aviation airport $p$ to abortion clinic $c$ by county $q$ on mode $j$
\item $x_{q, c, j}$: flow from county $q$ to abortion clinic $c$ on mode $j$
\item $x_{q, c}$: total flow from county $q$ to abortion clinic $c$
\end{itemize}

\subsection{Maximum Flow Model}
\label{sec:max_flow}
The objective function that maximizes the total flow across the network is:
\begin{align}
\max \text{  }  \sum_{q \in Q}  \sum_{c \in C} x_{q,c}
\end{align}

The constraints in our model are described as follows: Eq. \eqref{eq:c1} ensures that the flow out of each county cannot be greater than the estimated demand of abortion for that county. Eq. \eqref{eq:c2} restricts the flow from an origin county to an origin airport and from an origin county to a clinic to be no more than the total volunteer vehicles available for that county. Eq. \eqref{eq:c3} ensures that the flow from destination airports to clinics be less than the available supply of volunteer drivers in the destination county. Eq. \eqref{eq:c4} says that the flow from origin to destination general aviation airports should be less than the total available seats on general aircraft available. Eq. \eqref{eq:c5} ensures that the flow from an origin commercial airport to a destination commercial airport be less than the total available flight seats for that pair of airports. Eq. \eqref{eq:c6} restricts the flow into a particular clinic to be no more than the clinic’s available capacity for the day. Eq. \eqref{eq:c7} is the budget constraint that constrains all transportation costs to be less than the total nonprofit’s budget for the day. There is no cost associated with transporting individuals on private vehicles or on general aircraft. Note that in Eqs. \eqref{eq:c3} -- \eqref{eq:c5} and Eq. \eqref{eq:c7} we account for whether patients are traveling with companions by reducing the available supply of volunteer drivers, pilots, and maximum budget by a half. For the time constraints, Eqs. \eqref{eq:t1} -- \eqref{eq:t7} ensure that each arc in the network is restricted by the corresponding airport access/egress, in-flight, and county to clinic maximum travel times. Eqs. \eqref{eq:f1} -- \eqref{eq:f5} are the flow balance constraints and Eq. \eqref{eq:int} ensure that the decision variables are non-negative integers.  

\begin{align}
\sum_{j \in J} \sum_{g \in G} x_{q, g, j} + \sum_{j \in J} \sum_{m \in M} x_{q, m, j} + \sum_{c \in C} \sum_{j \in J} x_{q, c, j} \notag \\ \leq p_q, \text{  }\forall \text{  }q \in Q  \label{eq:c1}
\end{align}
\begin{align}
\sum_{g \in G} x_{q, g, j = PV} + \sum_{m \in M} x_{q, m, j = PV} + \sum_{c \in C} x_{q, c, j = PV} \notag \\ \leq \frac{w \cdot n_q}{y + 1} \text{  }\forall \text{  }q \in Q \label{eq:c2}
\end{align}
\begin{align}
\sum_{q \in Q} \sum_{c \in C} \sum_{p \in P} b_{p, v} x_{q, p, c, j = PV} + \sum_{q \in Q} \sum_{c \in C}  \sum_{r \in R} b_{r, v} x_{q, r, c, j = PV} \notag \\ \leq \frac{w \cdot n_{v}}{y + 1}, \text{   } \forall \text{   } v \in V \label{eq:c3}
\end{align}
\begin{align}
\sum_{q \in Q} \sum_{g \in G}  \sum_{p \in P} x_{q, g, p} \leq \frac{z \cdot u}{y + 1} \label{eq:c4}
\end{align}
\begin{align}
\sum_{q \in Q} x_{q, m, r} \leq \frac{f_{m, r}}{y+1}, \text{  }\forall \text{  }m \in M, r \in R \label{eq:c5}
\end{align}
\begin{align}
\sum_{q \in Q} x_{q, c} \leq \lambda_c, \text{  }\forall \text{  }c \in C \label{eq:c6}
\end{align}
\begin{align}
\sum_{q \in Q} \sum_{m \in M} \sum_{r \in R} x_{q, m, r} a_{m, r} + o [ \sum_{q \in Q} \sum_{m \in M} x_{q, m, j = RH} t_{q, m, j = RH} \notag \\ +  \sum_{q \in Q} \sum_{g \in G} x_{q, g, j = RH} t_{q, g, j = RH} + \sum_{q \in Q} \sum_{c \in C} x_{q, c, j = RH} t_{q, c, j = RH} \notag \\ +  \sum_{q \in Q} \sum_{r \in R} \sum_{c \in C} x_{q, r, c, j = RH} t_{r, c, j = RH} \notag \\ + 
\sum_{q \in Q} \sum_{p \in P} \sum_{c \in C} x_{q, p, c, j = RH} t_{p, c, j = RH} ] \leq \frac{B}{y + 1} \label{eq:c7}
\end{align}
\begin{align}
x_{q, c, j} (t_{q, c, j} - T) \leq 0, \text{  }\forall \text{  }q \in Q, c \in C, j \in J \label{eq:t1}
\end{align}
\begin{align}
 x_{q, g, j} (t_{q, g, j} - \mathcal{A}) \leq 0, \text{  }\forall \text{  }q \in Q, g \in G, j \in J \label{eq:t2}
\end{align}
\begin{align}
x_{q, m, j} (t_{q, m, j} - \mathcal{A}) \leq 0, \text{  }\forall \text{  }q \in Q, m \in M, j \in J \label{eq:t3}
\end{align}
\begin{align}
x_{q, r, c, j} (t_{r, c, j} - \mathcal{A}) \leq 0, \text{  }\forall \text{  }q \in Q, r \in R, c \in C, j \in J \label{eq:t4} 
\end{align}
\begin{align}
x_{q, p, c, j} (t_{p, c, j} - \mathcal{A}) \leq 0, \text{  }\forall \text{  }q \in Q, p \in P, c \in C, j \in J
\label{eq:t5} 
\end{align}
\begin{align}
x_{q, m, r} (t_{m, r} - \mathcal{F}) \leq 0, \text{  }\forall \text{  }q \in Q, m \in M, r \in R \label{eq:t6} 
\end{align}
\begin{align}
x_{q, g, p} (t_{g, p} - \mathcal{F}) \leq 0, \text{  }\forall \text{  }q \in Q, g \in G, p \in P \label{eq:t7} 
\end{align}
\begin{align}
\sum_{j \in J} x_{q, m, j} =  \sum_{r \in R} x_{q, m, r}, \text{  }\forall \text{  }q \in Q, m \in M
\label{eq:f1} 
\end{align}
\begin{align}
\sum_{j \in J} x_{q, g, j} =  \sum_{p \in P} x_{q, g, p}, \text{  }\forall \text{  }q \in Q, g \in G
\label{eq:f2}
\end{align}
\begin{align}
\sum_{m \in M} x_{q, m, r} =  \sum_{c \in C} \sum_{j \in J} x_{q, r, c, j}, \text{  }\forall \text{  }q \in Q, r \in R
\label{eq:f3}
\end{align}
\begin{align}
\sum_{g \in G} x_{q, g, p} =  \sum_{c \in C} \sum_{j \in J} x_{q, p, c}, \text{  }\forall \text{  }q \in Q, p \in P
\label{eq:f4}
\end{align}
\begin{align}
\sum_{j \in J} \sum_{r \in R} x_{q, r, c, j} + \sum_{j \in J} \sum_{p \in P} x_{q, p, c, j} + \sum_{j \in J} x_{q, c, j} = x_{q, c} \notag \\ \text{  }\forall \text{  }q \in Q, c \in C
\label{eq:f5}
\end{align}

\begin{align}
 x_{q, g, j}, x_{q, m, j}, x_{q, g, p}, x_{q, m, r}, x_{q, r, c, j}, x_{q, p, c, j}, x_{q, c, j}, \notag \\ x_{q, c} \in \mathbb{Z}^+ \text{  }\forall \text{  }q \in Q, c \in C, g \in G, m \in M, \notag \\ j \in J, r \in R, p \in P
\end{align}
\label{eq:int}

\subsection{Minimum Cost Model}
\label{sec:min_cost}
To formulate the minimum cost model, the objective function that minimizes the transportation cost across the network is:
\begin{align}
\min \sum_{q \in Q} \sum_{m \in M} \sum_{r \in R} x_{q, m, r} a_{m, r} + o [ \sum_{q \in Q} \sum_{m \in M} x_{q, m, j = RH} t_{q, m, j = RH} \notag \\ +  \sum_{q \in Q} \sum_{g \in G} x_{q, g, j = RH} t_{q, g, j = RH} + \sum_{q \in Q} \sum_{c \in C} x_{q, c, j = RH} t_{q, c, j = RH} \notag \\ +  \sum_{q \in Q} \sum_{r \in R} \sum_{c \in C} x_{q, r, c, j = RH} t_{r, c, j = RH} \notag \\ + 
\sum_{q \in Q} \sum_{p \in P} \sum_{c \in C} x_{q, p, c, j = RH} t_{p, c, j = RH}]
\end{align}

For the minimum cost model, we use the same constraints as described in Section \ref{sec:max_flow}. However, we make the following two modifications to the maximum flow model constraints: First, we modify Eq. \eqref{eq:c1} to ensure that demand is completely satisfied: 
\begin{align}
\sum_{j \in J} \sum_{g \in G} x_{q, g, j} + \sum_{j \in J} \sum_{m \in M} x_{q, m, j} + \sum_{c \in C} \sum_{j \in J} x_{q, c, j} \notag \\ = p_q, \text{  }\forall \text{  }q \in Q  \label{eq:mc1}
\end{align}
Second, we eliminate Eq. \eqref{eq:c7} since the budget is no longer a constraint, but is a quantity minimized in the objective function.

\subsection{Front End Interface}
\label{sec:interface}
The second component of the RHEA methodology is the interactive, front end interface that allows the nonprofit user to vary input resource estimate parameters (e.g. maximum budget, volunteer drivers, and volunteer pilots), travel time constraints, and clinic openings and observe the effect on total network throughput, transportation cost, modal distributions, and abortion demand satisfaction. The user interface is a website created using Anvil, which provides an interactive front end that is connected to the optimization model back end. Through a process called Anvil Uplink, Anvil allows for the connection of a Jupyter Notebook to its cloud-based server that is hosted on AWS in London. With this back end connected, an appropriate front end can be built on the Anvil platform. Once it is built, Anvil provides the capability to turn the web application into a direct website that can be accessed from any computer. Prior to presenting the front end display and model outputs in Section \ref{sec:results}, we discuss the model implementation in Section \ref{sec:scope_data} next.

\section{Implementation}
\label{sec:scope_data}
In this section, we describe our model implementation and the geographic scope and data required to parameterize our model. Recall that our model is implemented on a state level: That is, we allow the user to input a particular origin and destination state over which to determine the maximum throughput, demand satisfaction, and minimum cost. Following the selection of the origin and destination states, we then instantiate the sets of origin and destination airports, based on the methodology described in Section \ref{sec:airports_geo}. Finally, the nonprofit user will input their resource estimates, travel time constraints, and clinic openings (Figure \ref{fig:chart1}).

\subsection{Selection of Airports and Geography}
\label{sec:airports_geo}
\begin{figure}[!htbp]
\centering
\includegraphics[width=1\columnwidth]
{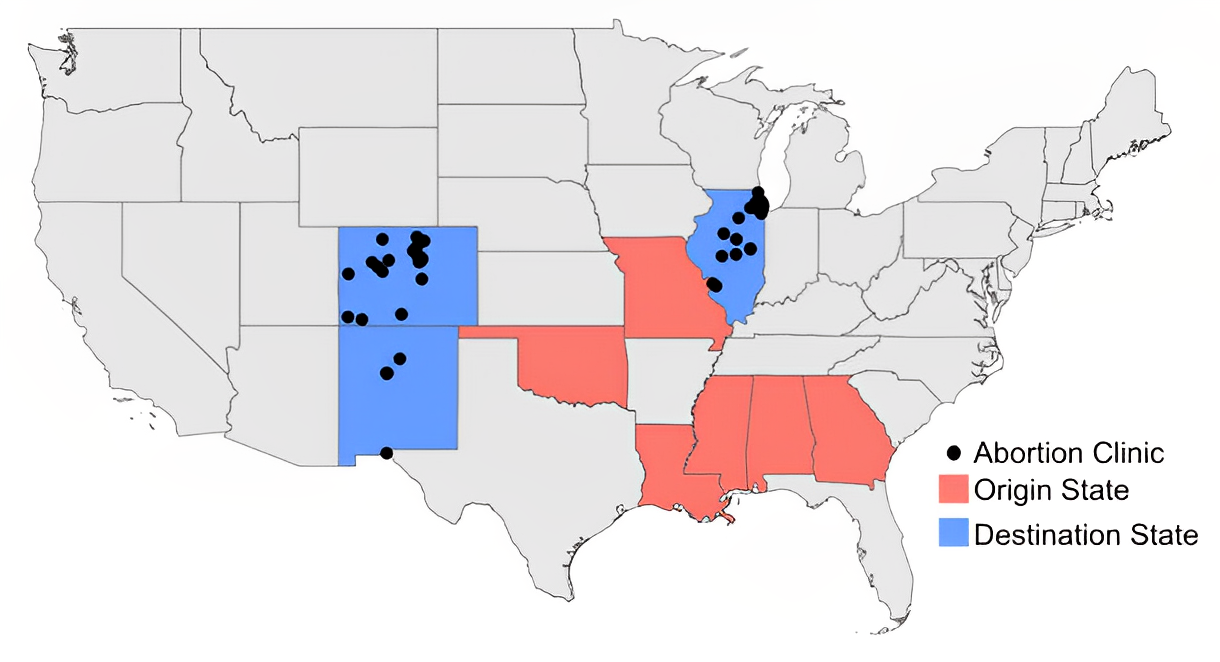}
\caption{Map of origin and destination states and abortion clinics.}
\label{fig:map}
\end{figure}

We first discuss the selection of geographic regions (defined by state boundaries) where women are likely to originate their trips, as well as the origin and destination airport sets. We employ the following search criterion to select the set of origin states: Based on \cite{guttmacher}, we find that the most restrictive states for abortion laws are located in the Southern region of the U.S. We then select those states who have a sizable population for us to model demand; our final selection of origin states includes Oklahoma, Missouri, Louisiana, Mississippi, Alabama, Georgia. The set of destination states to transport women to is selected based on each state's current abortion restriction status. We select those states that have the least abortion restrictions and are more centrally located to our origin states. The final selection of destination states includes Illinois, Colorado, and New Mexico. We discussed these selections of origin and destination states with our nonprofit partner, the Midwest Access Coalition, to validate these decisions for our model scope. These geographies are visualized in Figure \ref{fig:map}. 

For each origin and destination state, we determine the sets of commercial and general airports based on the following methodology: For origin airports corresponding to a particular origin state, we gather data on commercial ($M$) and general ($G$) airports located in the states: Kansas, Oklahoma, Texas, Louisiana, Arkansas, Missouri, Kentucky, Tennessee, Mississippi, Alabama, Georgia, Florida, South Carolina, and North Carolina. In doing so, we recognize that demand can be spread out across a state, so we allow for the most efficient transportation routes to airports depending on where demand is likely to originate from. As many abortion clinics are located near major cities with hub airports, we restrict the destination airports for commercial ($R$) and general ($P$) airports to those located in the destination state itself. Additionally, this selection is made to improve the user experience as it limits the number of destination counties containing a destination airport in which the user must input the number of volunteer drivers available. Future iterations of the model may allow for greater airport selection; however, for the purposes of the user experience, we limit our origin and destination airports based on the criterion above.

\subsection{Data Collection}
\label{sec:data_col}
In this section, we discuss the data used and methods to estimate the parameters listed in Section \ref{sec:parameters}. These parameters and data sources are summarized in Table \ref{tab:table1}.

\begin{table}[t]
\centering
\begin{tabular}{|l|c|c|}
\hline
\textbf{\thead{Set / \\ Parameter}} & \textbf{Description} & \textbf{Data Source}\\
\hline
\makecell{$p_q$} & \makecell{Demand for abortions \\ for women aged 15-44 \\ by county $q \in Q$} & \makecell{- 5-year American \\ Community Survey (ACS) \\ - Guttmacher Data Center} \\
\hline
\makecell{$M, G, P, R$} & \makecell{Set of commercial and \\ general airports} & \makecell{Bureau of Transportation \\ Statistics Master Coordinate \\ Aviation Support Table} \\
\hline
\makecell{$C$} & \makecell{Set of abortion clinics} & \makecell{Advancing New Standards \\ in Reproductive Health \\ Abortion Facility Database} \\
\hline
\makecell{$f_{m, r}, a_{m, r}$, \\ $t_{m, r}$} & \makecell{Commercial flight seats, \\ airfare, and travel time \\ from commercial airports \\ $m$ to $r$} & \makecell{Tequila Kiwi.com \\ Search API} \\
\hline
\makecell{$t_{q, g, j}, t_{q, m, j}$, \\ $t_{r, c, j}, t_{p, c, j},$ \\ $ t_{q, c, j}$} & \makecell{Surface travel times \\ between counties, clinics, \\ and airports by mode} & \makecell{Google Maps API} \\
\hline
\end{tabular}
\caption{Description of key sets, parameters, and data sources.}
\label{tab:table1}
\end{table}

\subsubsection{User-inputted parameters}
\label{sec:ui_params}
After the user has selected an selected origin and destination state, there are three subsequent types of user-inputted parameters. First, the user must input the clinics that are open for the destination state. To determine the baseline set of clinics from which the user can select the open clinics, we extract information for abortion facilities from the Advancing New Standards in Reproductive Health Abortion Facility Database \cite{ansirh}. We select only facilities that are open as of 2021 and are not telehealth-only clinics. As this data is not updated to reflect the most recent clinic openings and closures, we allow for the user to select the set of clinics that they know are open, from the original, baseline clinic data from the Abortion Facility Database. In doing so, we provide the nonprofit user with the flexibility to optimize operations based on real-time clinic closures.

Second, the nonprofit user selects their transportation supply and budget constraints (i.e. resource estimates), including: $u$ (volunteer pilots on standby), $B$ (total budget of the nonprofit for the day), $\lambda_c$ (number of individuals a clinic can service in a day), $n_q$ (volunteer driver supply at the origin state for an origin county $q$), $n_v$ (volunteer driver supply at the destination state for a destination county $v$) and $y$, a binary variable denoting whether the women to be transported are all traveling with a companion or not. We ensure that the supply of volunteer drivers in the origin and destination state is specific to the county; that is, women may only be transported by volunteer drivers who are located in the specific county in which they are either originating their travel. Additionally, in the destination state, women can only be transported to clinics by drivers who are located in the county in which their destination airport is located. In doing so, we acknowledge that there is a supply of volunteer drivers local to the county that can possibly transport women between counties and airports. This consideration also presents volunteers from driving long distances to pick up a patient and drop them off at their destination. 

Finally, the last type of user-inputted parameter are the travel time constraints, which include: $\mathcal{A}$ (maximum airport access and egress travel time), $\mathcal{F}$ (maximum in-flight time between origin and destination airports), and $T$ (maximum direct county to clinic travel time). These parameters ensure that the travel time for any leg of a route is constrained by an upper bound. To input these values, the nonprofit must have an understanding of how long they want each trip to take, based on their knowledge of volunteer and traveler preferences. This feature also allows volunteer pilots and drivers to avoid long travel if desired. 

\subsubsection{Transportation data collection}
As input data to our model, we collect commercial and general aviation data. First, we gather data on U.S. general and commercial airports from the Bureau of Transportation Statistics Master Coordinate Aviation Support Table \cite{bts}. Commercial aviation service characteristics are obtained using the Tequila by Kiwi.com Search API that extracts real-time flight data on an origin-destination airport basis \cite{kiwi}. For flight seat availability ($f_{m, r}$), we look at the total count of available seats from an origin commercial airport $m$ to a destination commercial airport $r$, booking three weeks in advance. To parameterize airfare ($a_{m, r}$) and travel time between origin and destination commercial airports ($t_{m, r}$), we take the average of airfare and travel time (which includes arrival and departure delay) associated with the available seats between an origin commercial airport $m$ and destination commercial airport $r$. For general aviation, there is no cost associated with transportation as the individual pilots incur the fuel and maintenance costs. To calculate travel time on general aircraft ($t_{g, p}$), we calculated the network distance between an origin general aviation airport $g$ and a destination general aviation airport $p$ using spatial packages in R and apply a speed of 130 mph. This speed was determined based on our discussions with Elevated Access, one of our nonprofit partners. We also set the general aircraft capacity ($z$) at four individuals per aircraft.

We estimate the surface travel time parameters ($t_{q, g, j}, t_{q, m, j}$, $t_{r, c, j}, t_{p, c, j}, t_{q, c, j}$) using the Google Maps API to calculate the shortest route between origins and destinations. For the options that individuals have in ground transportation modes, we limit the set to include only ride hail (i.e. Uber and Lyft) and volunteer drivers. We set the private vehicle capacity ($w$) at two individuals per vehicle. We exclude public transit (i.e. intercity bus, train) as an option for ground transportation to and from airports and directly from counties to clinics because few geographies we are considering have access via public transit. For surface transportation costs, there is only a cost associated with taking ride hail which we set at \$0.40 per minute based on standard estimates from Uber. 

\subsubsection{Demand estimation}
Finally, we estimate the demand for abortion for women aged 15-44 by county ($p_q$) in our origin states. For each origin state, we calculate the total population of women aged 15-44 in counties with a median income of less than 35,000, which we obtain from the 5-year American Community Survey \cite{acs}. As we are concerned with modeling one day's of operations for a nonprofit, we divide this total population by 365. To account for the likelihood that women may require an abortion, we scale our demand estimates by the observed demand data published by the Guttmacher Institute which provides the number of abortions per 1000 women aged 15–44 by the state of residence in 2020 \cite{gutt2}. Because we are using observed demand estimates, our demand estimation procedure may not reflect latent demand for abortion and thus may in fact be underestimated. Additionally, there is heterogeneity in need for reproductive healthcare due to individual socioeconomic circumstances, which our demand estimates do not capture precisely. However, we do emphasize that our demand estimation procedure can be modified as new data and statistics on abortion demand are published. We focus the contributions of this work on the optimization model framework and web interface development.

\section{Results}
\label{sec:results}
In this section, we present the results from the RHEA optimization model and front end interface. To illustrate the results, we simulate a hypothetical scenario for a nonprofit interested in serving demand for abortions for women in Missouri being transported to clinics in Illinois. To solve the maximum flow and minimum cost optimization models, we use the Gurobi Optimizer with a Python API. The optimization model is formulated as a pure-integer linear programming problem, and we thus require no non-linear approximation techniques and are guaranteed for the algorithm to provide us with an optimal solution. In solving the optimization, Gurobi uses a branch-and-bound technique as this is a pure integer problem.

\begin{figure*}[!htbp]
\centering
\includegraphics[width=1.7\columnwidth]
{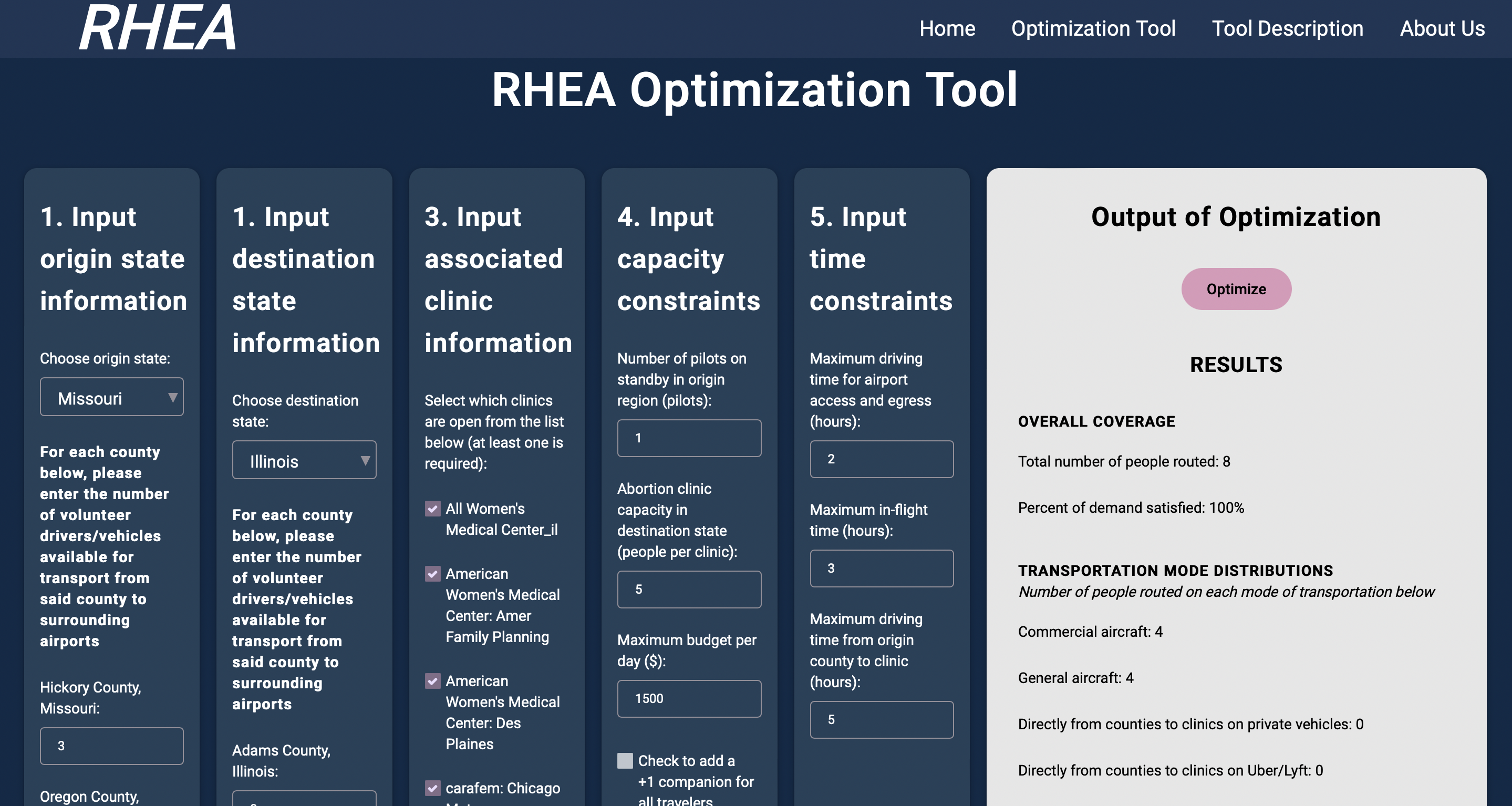}
\caption{RHEA front end interface with inputted parameters and optimization results.}
\label{fig:anvil_maxflow}
\end{figure*}

In the RHEA website displayed in Figure \ref{fig:anvil_maxflow}, the user is asked to input their origin and destination state, followed by the origin and destination volunteer drivers, clinic openings, capacity (i.e. budget, pilots, abortion clinics), and time constraints for a given day, which is described in Sections \ref{sec:parameters} and \ref{sec:ui_params}. To illustrate our results, we select the following input parameters for origin state Missouri and destination state Illinois: three volunteer drivers available in each of the origin counties in Missouri and 10 volunteer drivers in Cook County in Illinois. We only select Cook County in order to simulate a scenario where all the nonprofit's volunteer drivers are located in the region containing Chicago, which features a high density of clinics and large commercial hub airports (O'Hare and Midway International Airports). We also select the following clinics to be open in Illinois: All Women's Medical Center in Illinois, American Women's Medical Center (Amer Family Planning and Des Plaines), and carefem (Chicago Metro), which are all located in the Chicago area. The capacity constraints are: 1 pilot available on standby (with an aircraft capacity of four individuals), give abortions per day for each of the clinics, \$1500 maximum budget for the day, and no companions for all women. The time constraints are selected as two hours for maximum airport access and egress, three hours for in-flight time, and five hours for direct county to clinic time. 

\begin{figure}[!htbp]
\centering
\includegraphics[width=0.89\columnwidth]
{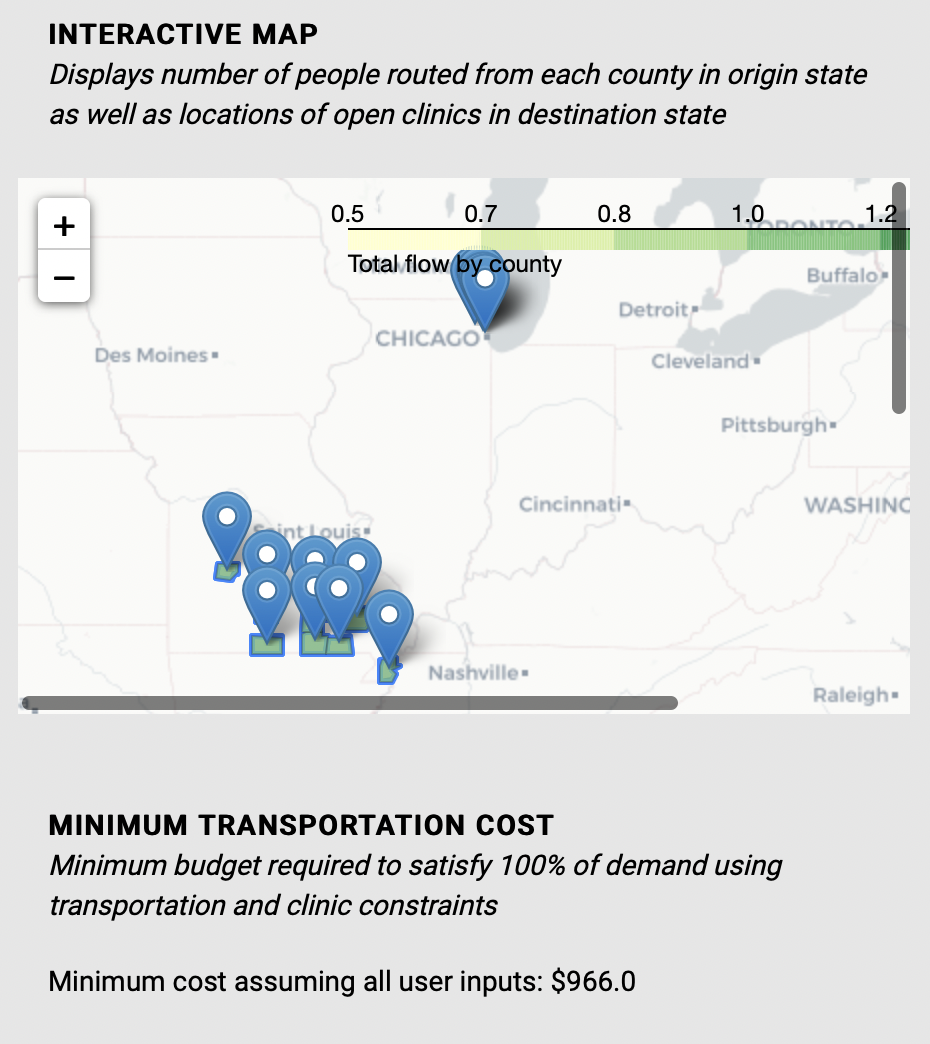}
\caption{Map of total women transported by county in Missouri and selected open clinics in Illinois, as well as the minimum transportation cost.}
\label{fig:spatial}
\end{figure}

Figure \ref{fig:anvil_maxflow} shows the inputted parameters and selected outputs of the optimization. The results show that in total, 8 women were able to be transported from Missouri to Illinois for one day, which corresponds to 100\% of demand satisfied. In Figure \ref{fig:spatial}, we display the total number of women transported from the Missouri counties to clinics, as well as the open clinics in Illinois. Based on our outputted statistics, we have that four women were routed on commercial aircraft, four were routed on general aircraft, and 0 women were routed directly from their origin county to a clinic. Because we inputted one pilot on standby, the solution uses all available capacity from general aircraft and then purchases commercial itineraries. The time constraint restricts direct county to clinic travel time to be less than five hours, which leads to no women routed directly from their origin county to a clinic. 

\begin{figure}[!htbp]
\centering
\includegraphics[width=0.795\columnwidth]
{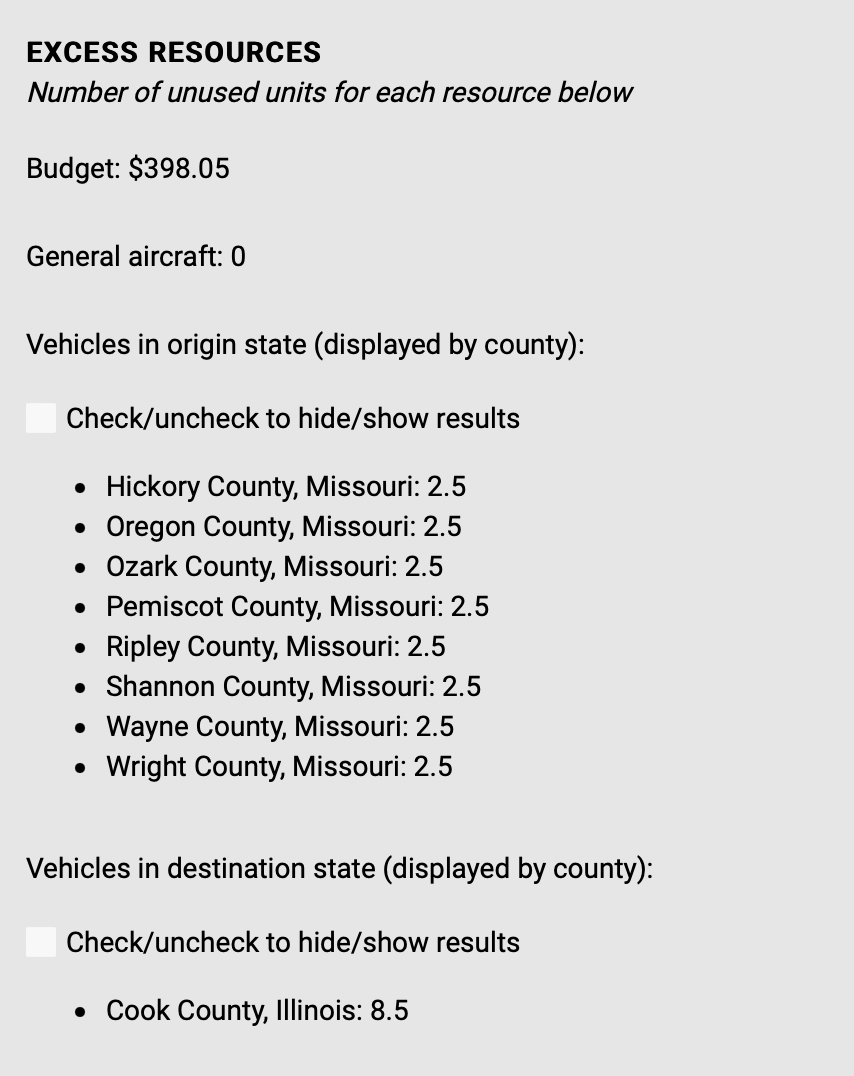}
\caption{Excess nonprofit resources for budget, general aircraft, and origin and destination volunteer drivers.}
\label{fig:excess}
\end{figure}

Additionally, we output the excess resources (i.e. budget, pilots, and drivers) to determine how much these resources were unused in the optimal solution, shown in Figure \ref{fig:excess}. For our example, we find that all general aircraft were used, as the slack for the general aircraft constraint is 0. The budget slack shows that \$398.05 were unused; this implies that if we were to re-run the optimization with \$398.05 less as the maximum budget, we would achieve the same optimal solution. The excess origin volunteer drivers to transport women between counties to airports shows that in all counties, only one vehicle was used, with each vehicle being half-full (as we assumed a capacity of two individuals per vehicle). With the destination volunteer drivers to transport women from airports to clinics, eight vehicles were unused, while one vehicle was completely full and the other was half-full. In our example, since the volunteer drivers were all located in Cook County in Illinois (which contains Illinois' largest commercial hub airports), three out of four individuals routed on commercial aircraft were transported to clinics by these volunteer drivers. The other five individuals who were transported from airports to clinics took ride hail (i.e. Uber/Lyft) from the destination airport to abortion clinic. 

It is important to note that the maximum flow model is not guaranteed to output the minimum cost solution, as the objective is to maximize total flow only, subject to a maximum budget constraint. To this end, our model and interface allows the nonprofit user to also understand the minimum amount of budget to be allocated to serve demand for abortions in the origin state, based on their existing constraints. In Figure \ref{fig:spatial}, the nonprofit user finds that \$966 is the minimum amount of funding required to satisfy abortion demand in Missouri, assuming the values of the input parameters originally selected. By displaying the minimum transportation cost solution, nonprofit users can understand how to best allocate resources to most effectively serve demand, while not wasting money or volunteer hours.

The results show that based on the nonprofit's constraints, the nonprofit may be able to serve a large portion of abortion demand. For planning and resource allocation purposes, the nonprofit can vary the input parameters and observe the effect on network throughput, demand satisfaction, modal distributions, and transportation cost. As such, RHEA provides a framework for the nonprofit user to input different combinations of resource estimates, travel time constraints, and clinic openings based on evolving traveler and volunteer preferences and dynamic supply restrictions.

\section{Concluding Remarks}
\label{sec:conclusion}
RHEA allows abortion access nonprofits to strategically allocate their resources (e.g. volunteer drivers, pilots, and monetary budget) under traveler and volunteer preferences (e.g. travel time constraints) and supply restrictions (e.g. clinic openings) to ensure that low-income, underserved women in the U.S. receive access to adequate reproductive healthcare. Using RHEA, abortion access nonprofits can understand how to best allocate their limited resources to serve demand for abortions. In doing so, we seek to alleviate healthcare inequity gaps so that underserved women have access to reproductive healthcare.

The RHEA optimization model and front end interface provides a framework that allows for further expansion. For our future work, we expand the set of origin and destination geographies. Additionally, instead of routing women on a state-level basis, we would remove these state boundaries and route women from an origin state to any possible destination clinic. As new data on abortion rates and statistics becomes publicly available, we plan to refine our demand estimation procedure to more precisely model demand patterns for abortions. Finally, we plan to integrate the option of providing abortion pills and telehealth services more explicitly into the model and front end interface.


\label{sec:future work}
\section*{Acknowledgments}
The team would like to thank Siddharth Deliwala and Dr. Jan Van Der Speigel for their continued support and guidance for this project. We would also like to thank the Advancing New Standards in Reproductive Health Abortion Facility Database from the University of California, San Francisco for providing us with abortion facility data. Finally, the team would also like to thank our partners: the Midwest Access Coalition, Elevated Access, and the Steel City Access Network for their continued support and input.

\bibliographystyle{IEEEtran}
\bibliography{reference}

\end{document}